\documentclass[conference]{IEEEtran}
\usepackage{cite}
\usepackage{amsmath,amssymb,amsfonts}
\usepackage[linesnumbered,ruled,vlined]{algorithm2e}
\usepackage{graphicx}
\usepackage{textcomp}
\usepackage{xcolor}
\usepackage{multirow}
\usepackage{moresize}

\usepackage{booktabs}

\usepackage{amsfonts}

\def\BibTeX{{\rm B\kern-.05em{\sc i\kern-.025em b}\kern-.08em
    T\kern-.1667em\lower.7ex\hbox{E}\kern-.125emX}}

\begin{document}

\title{Antenna Array Structures for Enhanced Cluster Index Modulation
\thanks{This work was supported by the Scientific and Technological Research Council of Turkey (TUBITAK) under grant no 218E035.}
\vspace{-2.5ex}
}

\author{Mahmoud~Raeisi$^*$, 
        Asil~Koc$^{\bullet}$, 
        Ibrahim~Yildirim$^{*,\star}$, 
        Ertugrul~Basar$^*$, 
        Tho Le-Ngoc$^{\bullet}$
        \\
        $^*$CoreLab, Department of Electrical and Electronics Engineering, Koc University, Sariyer 34450, Istanbul, Turkey
        \\
        $^{\bullet}$Department of Electrical and Computer Engineering, McGill University, Montreal, QC, Canada \\
        $^\star$Faculty of Electrical and Electronics Engineering, Istanbul Technical University, Sariyer 34469, Istanbul, Turkey\\
        Email: mraeisi19@ku.edu.tr, asil.koc@mail.mcgill.ca, yildirimib@itu.edu.tr,\\ ebasar@ku.edu.tr, tho.le-ngoc@mcgill.ca
\vspace{-3ex}
}

\maketitle

\begin{abstract}
This paper investigates the effect of various antenna array structures, i.e., uniform linear array (ULA), uniform rectangular array (URA), uniform circular array (UCA), and concentric circular array (CCA), on cluster index modulation (CIM) enabled massive multiple-input multiple-output (mMIMO) millimeter-wave (mmWave) communications systems. As the CIM technique indexes spatial clusters to convey additional information bits, the different radiation characteristics caused by different array structures can significantly affect system performance. By analyzing the effects of array characteristics such as radiation pattern, array directivity, half-power beam width (HPBW), and radiation side lobes on bit error rate (BER) performance, we reveal that URA achieves better error performance than its counterparts in a CIM-enabled mmWave system. We demonstrate that narrower beams alone cannot guarantee better BER performance in a CIM-based system. Instead, other radiation characteristics, especially radiation side lobes, can significantly influence system performance by entailing extra interference in the non-intended directions. Illustrative results show that URA owes its superiority to its lower side lobes. We also propose an algorithm to implement fixed phase shifters (FPS) as a hardware-efficient (HE) analog network structure (beamformer/combiner) to reduce cost and energy consumption in mmWave systems and investigate the effect of a non-ideal analog network on the  BER performance for different array structures. It is demonstrated that HE systems with a few FPSs can achieve similar BER performance compared to the optimum (OP) analog network structure.

\end{abstract}

\begin{IEEEkeywords}
Index modulation, mmWave, massive MIMO, array geometry, analog beamforming.
\vspace{-2ex}
\end{IEEEkeywords}

\section{Introduction}\vspace{-1ex}
\IEEEPARstart{M}{illimeter-wave} (mmWave) communications is a promising technology for future wireless communications due to its large spectrum availability, high bandwidth, and high data rate that can respond to the stringent demands of 6G communications. However, there are some challenges that mmWave communication encounters, e.g., severe path loss and attenuation, costly hardware equipment, and energy consumption. By exploiting the short wavelength of mmWave signals, we can pack large antenna arrays and utilize them in practical massive multiple-input multiple-output (mMIMO) communications systems. Large antenna arrays in mmWave systems compensate for the severe path loss by forming high directivity beams \cite{wang2020joint, wang2020intelligent}. Fully-digital beamforming (FDBF) is widely considered in conventional MIMO communications systems; however, it requires a power-hungry and expensive RF chain per antenna. Likewise, as the antenna array size increases, FDBF is not feasible for mMIMO systems due to the extremely high hardware cost/complexity \cite{koc2022full}. Analog beamforming (ABF) has recently received significant attention in the literature to be adopted in mmWave mMIMO communications systems since it needs only a single RF chain to perform the beamforming for a large antenna array \cite{koc2022full}. Nonetheless, implementing high-precision/coarsely-quantized\footnote{High-precision phase shifters are assumed in the literature to obtain near-optimal performance; however, it might be impractical. On the other hand, practical phase shifters have coarsely quantized phases; but they still should be implemented in large numbers to support mMIMO \cite{yu2018hardware}.} single phase shifters (SPS) per each RF chain-antenna pair is still needed in the analog network structure, which entails a costly and high energy consumption drawback in mmWave systems. A fixed phase shifter (FPS) structure is proposed in \cite{yu2018hardware} to reduce the number of phase shifters with quantized phases. It is shown that only a few phase shifters are required to implement the FPS structure in the analog network.

Index modulation (IM) is a promising technique to use in mmWave communications due to its ability to decrease the cost and energy consumption by decreasing the number of radio-frequency (RF) chains and increasing the spectral efficiency by conveying additional information bits \cite{basar2016index, basar2017index, jiang2021generalized}. The spatial scattering modulation (SSM) method, which provides additional information transmission by indexing orthogonal paths, is investigated in \cite{ding2017spatial, tu2018generalized, jiang2021generalized}. However, SSM might be im- practical for most mmWave environments due to the orthogonality assumption among paths. More recently, the authors of this article proposed cluster index modulation (CIM) in \cite{raeisi2022cluster} for mmWave mMIMO communications systems to enhance system performance by transmitting additional bits through indexing spatial clusters in the sparse mmWave environment. It has been demonstrated in \cite{raeisi2022cluster} that under practical circumstances of non-orthogonal paths, CIM performs better than SSM in terms of bit error rate (BER). 

In a sparse mmWave environment, clusters are typically located at large distances from each other, making them more suitable for indexing than paths, e.g., SSM scheme. The larger difference in physical distance results in less correlation among indexed entities, i.e., clusters in the CIM scheme, enhancing error performance. Nevertheless, interference from non-intended indexed clusters, potentially located far from the target cluster, is still expected at the receiver. Since inter-cluster interference is the primary source of interference in the CIM scheme, reducing such interference is of great importance. Accordingly, it is essential to form narrow beams in the steering direction and reduce side lobes in the other spatial angles to improve the error performance. Most of the works in the literature investigated the effect of different array structures on spectral/energy efficiency and/or data rate \cite{mahmood2021energy, mahmood2021massive, mahmood20202d, 8108586, liu2020machine, 9070133,7903703}. It is indicated that various array structures show different behaviors under different communication systems; hence, the results of a study with a specific communication method cannot be extended here. Besides, the effect of different antenna arrays on BER is less studied in the literature. The aforementioned gaps motivate us to have a more detailed look at the behavior of newly introduced CIM-enabled mmWave communication systems with various antenna array structures that are more adopted in the literature \cite{mahmood2021energy, mahmood2021massive, mahmood20202d, 8108586, liu2020machine, 9070133,7903703}.

In this paper, we investigate the effect of various antenna array structures, i.e., uniform linear array (ULA), uniform rectangular array (URA), uniform circular array (UCA), and concentric circular array (CCA), which are more considered in the literature \cite{mahmood2021energy, mahmood2021massive, mahmood20202d, 8108586, liu2020machine, 9070133,7903703}, on the error performance of the CIM-enabled mMIMO mmWave communications systems. More specifically, we study the radiation pattern characteristics of different antenna arrays and their effect on the system BER performance. We show that an efficient radiation pattern significantly influences the BER performance by reducing inter-cluster interference among the indexed clusters. We also reveal that a hardware-efficient (HE) analog network (beamformer/combiner) with a few FPSs can perform near the ideal analog networks when the studied antenna arrays are adopted. Adopting the HE structure reduces the cost and energy consumption significantly. To the best of our knowledge, the performance of different array structures on SSM/CIM-enabled mmWave communications systems with FPS analog networks has not been investigated in the existing literature.

The rest of this paper is organized as follows. Section \ref{Sec:System Model} illustrates the system model. Analog beamforming with different antenna array structures are discussed in Section \ref{Sec: Different Antenna Array Structures}.  Illustrative results are presented in Section \ref{Sec: Illustrative Results}, and Section \ref{Sec: Conclusion} concludes this work.

\vspace{-2ex}

\section{System Model}\label{Sec:System Model}
\vspace{-1ex}

\begin{figure}
    \centering
    \includegraphics[scale = 0.3]{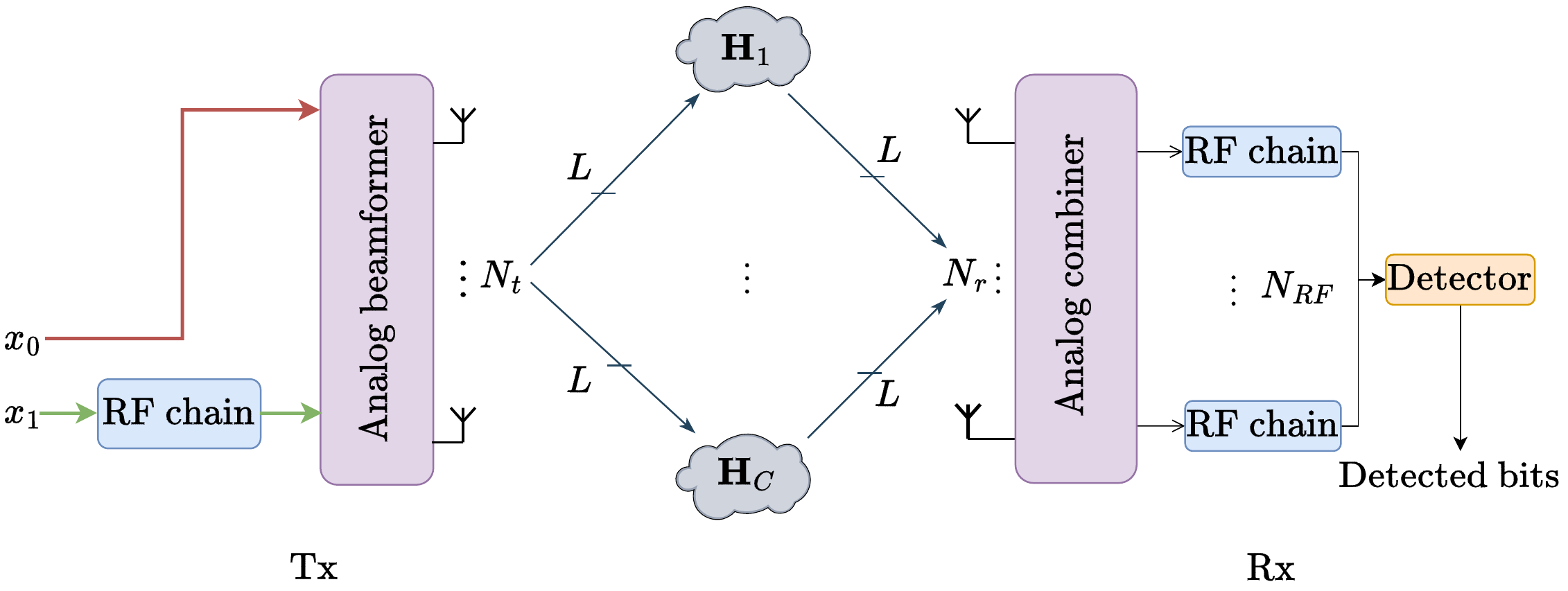}
    \caption{System model.}
    \label{fig:System Model} 
    \vspace{-4ex}
\end{figure}

This section presents the system model of the considered mMIMO system with ABF in a sparse mmWave environment, as illustrated in Fig. \ref{fig:System Model}. The number of antenna elements at the transmitter (Tx) and receiver (Rx) is assumed to be $N_t$ and $N_r$, respectively. For ease of analysis, we consider a point-to-point scenario in which only one RF chain is adopted at the Tx\footnote{This system model can be generalized to a multi-user/stream scenario with $K$ users/streams; hence, at least $K$ RF chains are needed at the Tx.}. By considering mMIMO at both terminals as a back-haul link, there are $N_{RF}$ RF chains at the Rx ($N_{RF} \ll N_r$) to form the receive beams toward the indexed clusters. Thanks to the CIM scheme, the Tx conveys an extra stream denoted by $x_0$ as spatial domain information without using an extra RF chain \cite{wang2019towards}. Due to the vulnerability of the mmWave signal to the blockage, we assume there is no line-of-sight (LoS) link between the Tx and the Rx \cite{wang2020intelligent}. Consequently, the transmission between them relies on non-line-of-sight (NLoS) channels, i.e., $\mathbf{H}_i \in \mathbb{C}^{N_r \times N_t}$, $i = 1, \dots, C$, where $C$ is the total number of clusters in the sparse environment. Besides, there are $L$ paths within each cluster.

\vspace{-2ex}

\subsection{Channel model}
\vspace{-1ex}
We consider Saleh-Valenzuela (SV) channel model as follows \cite{wang2020joint, wang2020intelligent, koc2022full, ying2020gmd, raeisi2022cluster}:
\begin{equation}\label{channel model}
\begin{split}
    \mathbf{H} = 
    & \sqrt{\frac{N_t N_r}{C L}}\sum_{c = 1}^{C} \sum_{l = 1}^{L} \alpha_{c,l} \mathbf{a}_{r}(\varphi_{c,l}^{r},\vartheta_{c,l}^{r}) \mathbf{a}_t^H(\varphi_{c,l}^t,\vartheta_{c,l}^t),
\end{split}
\end{equation}
where $\alpha_{c,l}$ is the NLoS complex channel gain of the $l$th path in the $c$th cluster, $\mathbf{a}_t(.)$ and $\mathbf{a}_r(.)$ are transmit and receive array response vectors\footnote{Section \ref{Sec: Different Antenna Array Structures} defines the array response vector for each array structure.}, respectively, $\varphi_{c,l}^t$ and $\vartheta_{c,l}^t$ are the azimuth and elevation angle-of-departure (AoD) at the Tx, respectively, $\varphi_{c,l}^r$ and $\vartheta_{c,l}^r$ are the azimuth and elevation angle-of-arrival (AoA) at the Rx, respectively. $\alpha_{c,l}$ follows complex Gaussian distribution with zero-mean and variance $\sigma^2$, i.e., $\alpha_{c,l} \sim \mathcal{C}\mathcal{N}(0,\sigma^2)$, wherein $\sigma^2 = 10^{-0.1 PL(d)}$. Notice that the path loss between the Tx and the Rx is shown by $PL(d)$, and $d$ is the distance between the terminals. Each cluster is specified by its azimuth and elevation mean angles at Tx (Rx), i.e., $\varphi_m^t  (\varphi_m^r$) and $\vartheta_m^t  (\vartheta_m^r$), which are uniformly distributed\footnote{A uniform distribution with parameters $\beta_1$ and $\beta_2$ is shown as $\mathcal{U}(\beta_1,\beta_2)$.}. Within each cluster, the paths are spread according to Laplacian distribution with associated azimuth/elevation mean angles and the angular spread of azimuth/elevation domains, i.e., $\sigma_{\varphi_m^t}, \sigma_{\vartheta_m^t}, \sigma_{\varphi_m^r},  \sigma_{\vartheta_m^r}$ \cite{liu2020machine, el2014spatially}.
\vspace{-1ex}

\subsection{Signal model and cluster index modulation}
\vspace{-1ex}

The proposed system model transmits two information streams; $x_0$ is the CIM symbol including $\log_2(B)$ bits, and $x_1$ is the $M$-ary symbol including $\log_2(M)$ bits. Here $M$ is the $M$-ary constellation order while $B$ is the CIM codebook order. In other words, $B$ ($B \leq N_{RF}$) is the number of indexed clusters. Accordingly, the spectral efficiency is $\eta = \log_2(B) + \log_2(M)$ bits per channel use (bpcu). 
The received signal $\mathbf{y} \in \mathbb{C}^{N_r \times 1}$ at the receive antenna array terminal is 
\vspace{-1ex}
\begin{equation}
    \mathbf{y} = \sqrt{P} G_t G_r \mathbf{H}\mathbf{f}s + \mathbf{n},
    \vspace{-1ex}
\end{equation}
where $\mathbf{f} \in \mathbb{C}^{N_t \times 1}$ is the analog beamformer vector, $s$ is the modulated symbol using the $M$-ary constellation, $\mathbf{n} \in \mathbb{C}^{N_r \times 1} \sim \mathcal{C}\mathcal{N}(0, \sigma_N^2)$ is the additive noise component at the Rx with variance $\sigma_N^2$, $P$ is the transmit power, and $G_t$ ($G_r$) \cite{wang2020joint, ning2021terahertz} is the Tx (Rx) antenna array gain.
The Rx forms multiple receive beams toward the indexed clusters by adopting multiple analog combiners in order to resolve the spatial CIM symbol, i.e., $x_0$, \cite{ding2017spatial,raeisi2022cluster}. Therefore, the received signal $\mathbf{z} \in \mathbb{C}^{B \times 1}$ after processing by analog combiner block is obtained as 
\vspace{-1ex}
\begin{equation}
    \mathbf{z} = \sqrt{P} G_t G_r \mathbf{W}^H \mathbf{H} \mathbf{f} s + \mathbf{W}^H \mathbf{n},
    \vspace{-1.3ex}
\end{equation}
where $\mathbf{W} = [\mathbf{w}_1, \dots, \mathbf{w}_B] \in \mathbb{C}^{N_r \times B}$ is the analog combiner matrix block, with $\mathbf{w}_i \in \mathbb{C}^{N_r \times 1}$ being the analog combiner vector associated with the $i$th ($i = 1, \dotsm B$) indexed cluster.
A maximum-likelihood detector is adopted to jointly resolve the CIM and $M$-ary symbols as follows:\vspace{-1ex}
\begin{equation}
    [\hat{c},\hat{s}] = \arg \min_{c,s} \big | \mathbf{z}(c) - \sqrt{P} G_t G_r \mathbf{w}_{c}^H \mathbf{H} \mathbf{f} s \big |^2, 
    \vspace{-1.3ex} 
\end{equation}
where the estimated CIM and $M$-ary symbols are shown as $\hat{c}$ and $\hat{s}$, respectively, and $\mathbf{z}(c)$ represents the $c$th element in the received signal vector $\mathbf{z}$.

\setlength{\textfloatsep}{0pt}
\begin{algorithm}[t]
\footnotesize
\caption{$\mathcal{B}$ and $\mathbf{W}$ construction algorithm.}\label{alg:CIM codebook construction}
\KwIn{$\mathbf{H}, \mathbf{a}(.)$}
\KwOut{$\mathcal{B}$, $\mathbf{W}$}
    Find the best effective path for each cluster using (\ref{eq:Best effective path index}).\\
Construct vector
$\mathcal{P} = \{ p_1, \dots, p_{C} \}.$\\
\For{$k \gets 1 \ to \ B$}{
    calculate $k$th codeword as $\mathbf{b}_k = \mathbf{f}_{b_k,p_{b_k}}$ via (\ref{eq:beamformer}) and (\ref{eq:CIM codeword}).\\ 
    calculate $k$th combiner as $\mathbf{w}_k = \mathbf{w}_{b_k,p_{b_k}}$ via (\ref{eq:combiner}) and (\ref{eq:CIM codeword}).\\
    Update $\mathcal{P}$:
    $\mathcal{P} = \mathcal{P} - \{b_k\}.$
}
Construct CIM codebook: $\mathcal{B} = \{ \mathbf{b}_{1}, \dots, \mathbf{b}_B \}.$\\
Construct analog combiner: $\mathbf{W} = [\mathbf{w}_{1}, \dots, \mathbf{w}_{B}]$.
\end{algorithm}

Analog beamformer $\mathbf{f}$ is selected from a pre-defined CIM codebook according to the CIM symbol $x_0$. Once the CIM codebook is constructed, the analog combiner matrix block can be constructed accordingly\footnote{In this paper, we assume that the channel state information (CSI) is perfectly known at the Tx and the Rx \cite{9685434}. Therefore, all path gains, AoDs, and AoAs associated with the paths/clusters are available at the Tx and Rx.}. In order to build a pre-defined codebook, each transmit array response vector associated with the corresponding indexed cluster is mapped to the corresponding CIM codeword. 
Hence, CIM codebook $\mathcal{B}$ includes $B$ codewords that specify the directions of indexed clusters. In each cluster, the best effective path is considered to steer the beam toward it. The best effective path in the $c$th cluster is defined as follows:
\vspace{-1.7ex}
\begin{equation}\label{eq:Best effective path index}
    p_c = \arg \max_{l = 1, \dots, L} |\mathbf{w}_{c,l}^H \mathbf{H} \mathbf{f}_{c,l}|^2, \ c = 1, \dots, C,
    \vspace{-1ex}
\end{equation}
where $\mathbf{f}_{c,l} \in \mathbb{C}^{N_t \times 1}$ and $\mathbf{w}_{c,l} \in \mathbb{C}^{N_r \times 1}$ are respectively the beamformer and combiner to steer the beam toward the $l$th path within the $c$th cluster, defined as follows:\vspace{-1ex}
\begin{equation}\label{eq:beamformer}
    \mathbf{f}_{c,l} = \mathbf{a}_t(\varphi_{c,l}^t, \vartheta_{c,l}^t), 
    \vspace{-1ex}
\end{equation}
\begin{equation}\label{eq:combiner}
    \mathbf{w}_{c,l} = \mathbf{a}_r(\varphi_{c,l}^r, \vartheta_{c,l}^r). \vspace{-1ex}
\end{equation}
Henceforth, we show the set of indexed clusters as $\mathcal{P} = \{ p_1, \dots, p_C \}$. Ultimately, each codeword of the CIM codebook is achieved consecutively as follows:\vspace{-1ex}
\begin{equation}\label{eq:CIM codeword}
    \mathbf{b} = \mathbf{f}_{b,p_{b}}, \ \ b = \arg \max\limits_{c = 1, \dots, C} |\mathbf{w}_{c,p_c}^H \mathbf{H} \mathbf{f}_{c,p_c}|^2,
    \vspace{-1.5ex}
\end{equation}
where $\mathbf{b} \in \mathbb{C}^{N_t \times 1}$ is the CIM codeword. Algorithm \ref{alg:CIM codebook construction} summarizes the instructions to construct $\mathcal{B}$ and $\mathbf{W}$.

\vspace{-1ex}

\section{Analog Beamforming with Different Array Structures}\label{Sec: Different Antenna Array Structures}

\vspace{-0.7ex}

This section describes different array structures and corresponding antenna array response vectors. Afterward, we propose implementing an HE structure for the analog network to be adopted in the practical mmWave mMIMO systems.

\begin{figure}
    \centering
    \includegraphics[scale = 0.4]{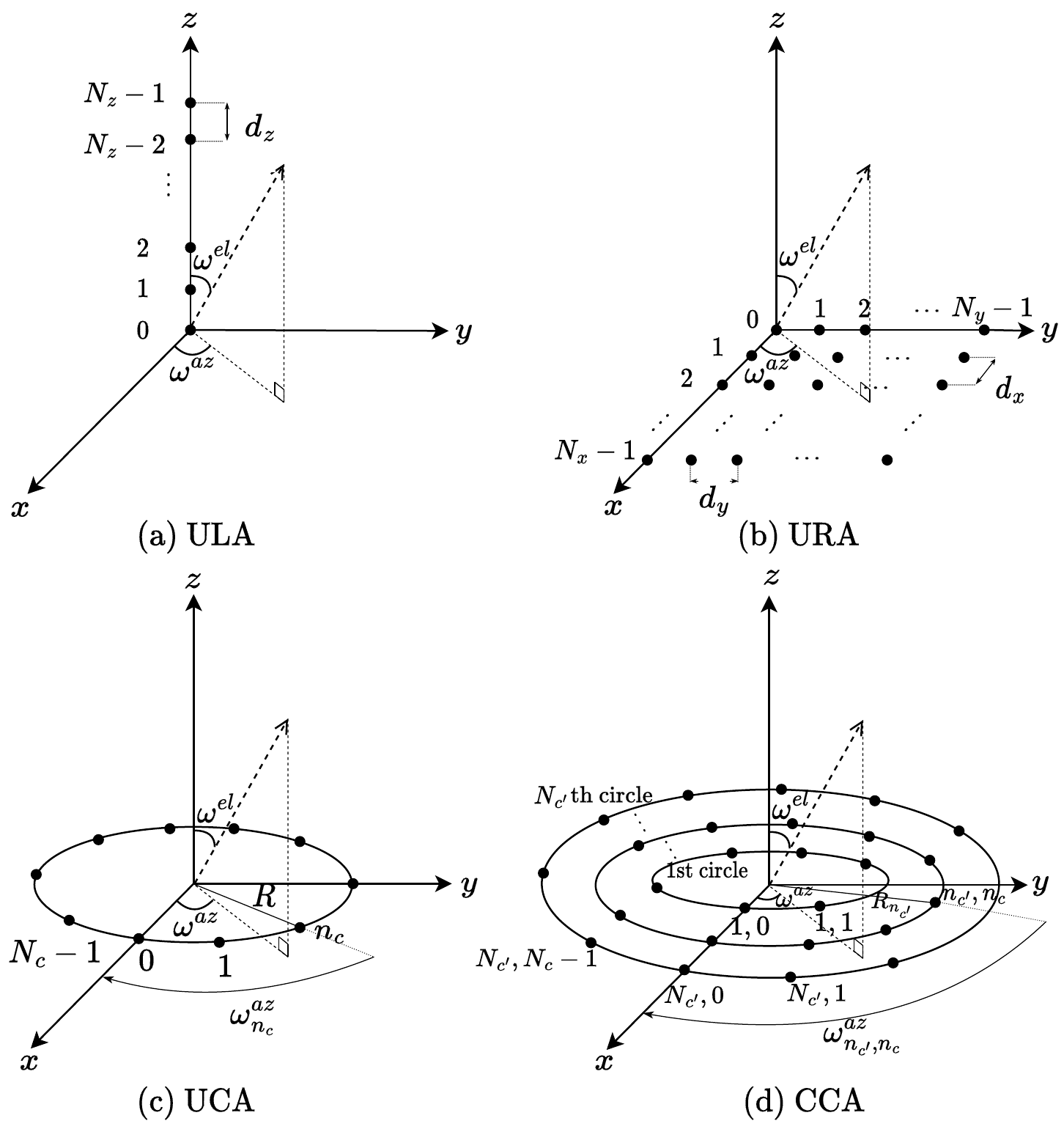}
    \caption{Different array geometries.}
    \label{fig:AA geometry} 
\end{figure}

\vspace{-1.5ex}
\subsection{Different antenna array structures}
\vspace{-0.6ex}

In \cite{mahmood2021energy, mahmood20202d}, it has been revealed that the proper array configuration effectively generates narrower beams. Here, besides narrower beams, reducing side lobes is also important.
The general expression for the array response vector is \cite{liu2020machine}
\vspace{-1ex}
\begin{equation}
    \mathbf{a}(\omega^{az}, \omega^{el}) = \frac{1}{\sqrt{N}}[e^{j\mathbf{k}^T \mathbf{p}_0}, e^{j\mathbf{k}^T \mathbf{p}_1}, ..., e^{j\mathbf{k}^T \mathbf{p}_{N-1}}]^T,
    \vspace{-1ex}
\end{equation}
where $\omega^{az} \in \{ \varphi_{c,l}^t, \varphi_{c,l}^r \}$($\omega^{el} \in \{ \vartheta_{c,l}^t, \vartheta_{c,l}^r \}$) is the corresponding angle of azimuth (elevation), $N \in \{N_t, N_r\}$ is the total number of antenna elements, $\mathbf{p}_n$ specifies the location of $n$th antenna element, and $\mathbf{k}$ is the wave-number defined as \cite{liu2020machine}
\vspace{-1ex}
\begin{equation*}
    \mathbf{k} = \frac{2 \pi}{\lambda}
    \begin{bmatrix}

        \sin (\omega^{el}) \cos (\omega^{az}), \

        \sin (\omega^{el}) \sin (\omega^{az}), \

        \cos (\omega^{el})

   \end{bmatrix}^T,
   \vspace{-0.8ex}
\end{equation*}
where $\lambda$ is the signal wavelength. The vector $\mathbf{p}_n$ is unique for each structure and specifies the array geometry. The corresponding $\mathbf{p}_n$ for ULA, URA, UCA, and CCA are given as follows, respectively \cite{liu2020machine}:
\vspace{-1ex}
\begin{equation}
    \mathbf{p}_{n_z} = 
    \begin{bmatrix}
        0,\
        0,\
        n_z d_z
    \end{bmatrix}^T,
\end{equation}
\begin{equation}
    \mathbf{p}_{n_x,n_y} = 
    \begin{bmatrix}
        n_x d_x,\
        n_y d_y,\
        0
    \end{bmatrix}^T,
\end{equation}
\begin{equation}
    \mathbf{p}_{n_c} = R
    \begin{bmatrix}
        \cos \omega^{az}_{n_c},\
        \sin \omega^{az}_{n_c},\
        0
    \end{bmatrix}^T,
\end{equation}
\begin{equation}
    \mathbf{p}_{n_{c'}, n_c} = R_{n_{c'}}
    \begin{bmatrix}
        \cos \omega^{az}_{n_{c'}, n_c},\
        \sin \omega^{az}_{n_{c'}, n_c},\
        0
    \end{bmatrix}^T,
    \vspace{-1ex}
\end{equation}
where $0 \leq n_x \leq N_x-1$, $0 \leq n_y \leq N_y-1$, $0 \leq n_z \leq N_z-1$ depict the location of antenna elements spanned on the $x$, $y$, $z$ directions, respectively, $0 \leq n_c \leq N_c-1$ is the location of antenna elements spanned on each circle of corresponding antenna structures; $1 \leq n_{c'} \leq N_{c'}$ is the $n_{c'}$th circle corresponding to CCA structure with radius $R_{n_{c'}}$; $R$ is the antenna array radius in UCA structure; and $d_x$, $d_y$ and $d_z$ are the space between two adjacent antenna elements along $x$, $y$, and $z$ direction, respectively. Furthermore, $\omega^{az}_{n_c}$ and  $\omega^{az}_{n_{c'},n_c}$ are depicted in Figs. \ref{fig:AA geometry}(c) and \ref{fig:AA geometry}(d), respectively, and are defined as $\omega^{az}_{n_{c'},n_c}  = \omega^{az}_{n_c} = \frac{2 \pi n_c}{N_c}$. The geometries of the aforementioned array structures are depicted in Fig. \ref{fig:AA geometry}.

\vspace{-1.8ex}
\subsection{Hardware-efficient analog network structure}
\begin{figure}
    \centering
    \includegraphics[scale = 0.4]{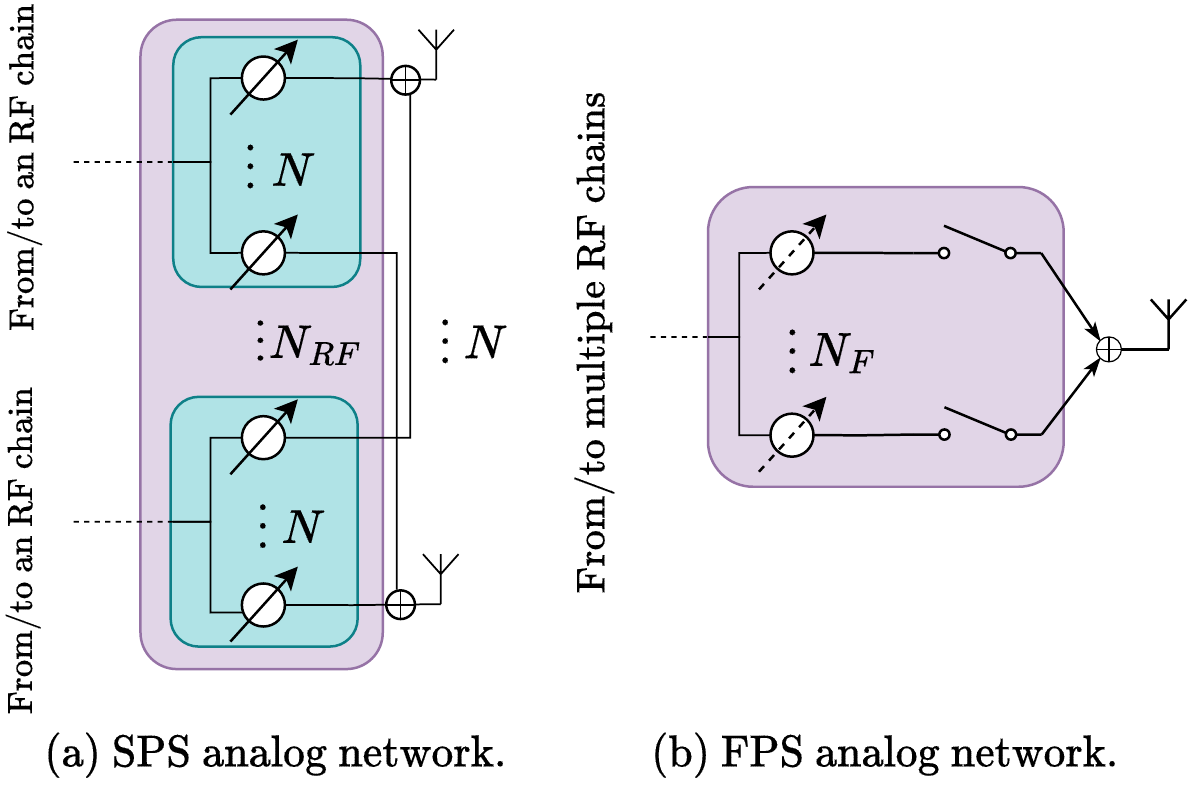}
    \caption{Analog network structures.}
    \label{fig:AnalogNetwork} 
\end{figure}

\vspace{-1ex}
In order to perform analog beamforming/combining to the intended direction, each RF chain is connected to the antenna array through a network of phase shifters at both terminals. In this paper, we investigate two structures for the analog network, i.e., near-optimal (OP) single phase shifters (SPS) structure with high-resolution phase shifters (as shown in Fig. \ref{fig:AnalogNetwork}(a)) and HE fixed phase shifters (FPS) structure with fixed and quantized phase shifters (as given in Fig. \ref{fig:AnalogNetwork}(b)). The SPS structure is widely used in the literature to achieve near-optimal performance \cite{wang2020joint, wang2020intelligent, koc2022full,  jiang2021generalized, ding2017spatial, raeisi2022cluster, liu2020machine}; however, it requires implementing a phase shifter per each RF chain-antenna pair. In order to implement the OP analog network, the phase shifters should have high-precision resolution and match with the continuous phases of antenna array response vectors. Hence, (\ref{eq:beamformer}) and (\ref{eq:combiner}) can be adopted as the beamformer and combiner, respectively.

In order to realize the HE analog network with a few numbers of phase shifters, the FPS structure is proposed in \cite{yu2018hardware}. As depicted in Fig. \ref{fig:AnalogNetwork}(b), the number of adopted phase shifters is not dependent on the other parameters, i.e., the number of antenna elements and RF chains. Instead, a network of adaptive switches is implemented to provide dynamic connections between phase shifters and antenna elements. It is worth mentioning that Fig. \ref{fig:AnalogNetwork}(b) focuses on a single pair of RF chain-antenna to describe the FPS structure clearly. Therefore, for each RF chain-antenna pair, deploying $N_F$ switches is required. Assume that we want to compose an analog beamformer/combiner to steer a beam toward the $l$th path of the $c$th cluster with antenna array response vector $\mathbf{a} = \frac{1}{\sqrt{N}} [e^{j\Theta_0}, e^{j\Theta_1}, \dots, e^{j\Theta_{N-1}}]$, where $\Theta_i$ is the continuous phase corresponding to the $i$th antenna. 
The signal flow passes through the $N_F$ fixed phase shifters and generates $N_F$ signals with different phases. To compose the signal with an approximation of a certain phase, a subset of $N_F$ signals can be combined by determining the statuses of $N_F$ adaptive switches. The quantized phase of each RF chain-antenna pair can be expressed as follows: \vspace{-1.5ex}
\begin{equation}\label{eq:quantized phase shift}
    \Omega = \mathbf{l}^T \mathbf{v},
    \vspace{-2ex}
\end{equation}
where $\mathbf{l}$ is the vector of switches status including $N_F$ binary elements, and $\mathbf{v}$ is the FPS vector including $N_F$ fixed phase shifters values. By adopting $N_F$ FPSs, obtaining $N_F-1$ bits resolution is possible if we define $\mathbf{v}$ as follows:
\begin{equation}
    \mathbf{v} = \frac{2 \pi}{2^{N_F-1}}[0, 2^0, 2^1, \dots, 2^{N_F-2}]^T, \ N_F \geq 2.
    \vspace{-1ex}
\end{equation}
Accordingly, adopting (\ref{eq:quantized phase shift}) allows the Tx/Rx to compose a desired quantized phase shift by combining appropriate FPSs. Therefore, the associated quantized beamformer (combiner) can be obtained as $\mathbf{f} = \frac{1}{\sqrt{N}} [e^{j\Omega_0}, e^{j\Omega_1}, \dots, e^{j\Omega_{N-1}}]$, where $\Omega_i$ is the quantized phase shift corresponding to the $i$th antenna element (quantized combiner can be composed similarly). Algorithm \ref{alg:Switch matrix construction} states how to determine switches status to compose quantized phase shift $\Omega$ associated with $\Theta$.

\begin{algorithm}[t]
\footnotesize
\caption{Switch vector composition algorithm.}\label{alg:Switch matrix construction}
\KwIn{$\Theta$}
\KwOut{$\mathbf{l}$}
Initialize $\mathbf{l}$ to zero vector (open all switches).\\ 
Wrap the phase shift $\Theta$ to $2 \pi$.\\
$t \leftarrow \Theta$.\\
$i \leftarrow \log_2{N_F} + 2$.\\
\While{$i \neq 0$}{
    \If{$\mathbf{v}(i) \leq t$}{
        $\mathbf{l}(i) \leftarrow 1$ (close the $i$th switch).\\
        $t \leftarrow t - \mathbf{v}(i)$
    }
    $i \leftarrow i - 1$.
}\vspace{-1ex}
\end{algorithm}

\begin{table}
\vspace{-3ex}
\footnotesize
    \centering
    \caption{Simulation setup parameters.}
    \vspace{-2ex}
    \begin{tabular}{ c c }
    \toprule 
    \textbf{Parameter}    & \textbf{value}  \\
    \toprule
    Tx location      & ($25$ m, $25$ m, $9$ m)   \\
    \midrule
    Rx location     & ($25$ m, $175$ m, $9$ m)    \\
    \midrule
    $\sigma_N^2$ \cite{wang2020intelligent, wang2020joint}     & $-90$ dBm    \\
    \midrule
    $G_t = G_r$   \cite{ning2021terahertz}    & 4 + 10 $\log_{10}(\sqrt{N})$    \\
    \midrule
    $a$, $b$, $\sigma_{\xi}$  \cite{akdeniz2014millimeter} & $72$, $2.92$, $8.7$ dB  \\
    \midrule
    $f_c$ \cite{akdeniz2014millimeter} & $28$ GHz  \\
    \midrule
    $d_x = d_y$ & $\lambda / 2$\\
    \bottomrule
\end{tabular}
\begin{tabular}{c c}
    \toprule
    \textbf{Parameter} & \textbf{value} \\
    \toprule
    $L$, $C$ \cite{liu2020machine, el2014spatially} & $10$, $8$ \\
    \midrule 
    $N$(Except URA) & $82$ \\
    \midrule
    $N$(Only URA) & $9 \times 9$ \\
    \midrule
    $R$ & $N \lambda / 4\pi$ \\
    \midrule
    $\varphi_m^t, \varphi_m^r$ & $\mathcal{U}[0,2\pi)$\\
    \midrule 
    $\vartheta_m^t, \vartheta_m^r$ & $\mathcal{U}[0,\pi)$\\
    \midrule 
    $\sigma_{\omega}$ \cite{liu2020machine, el2014spatially} & $7.5^\circ$ \\
    \bottomrule 
\end{tabular}
  \label{tab:System parameters} 
\end{table}

\vspace{-1.5ex}
\section{Illustrative Results}\label{Sec: Illustrative Results}
\vspace{-1ex}
This section presents computer simulation results in terms of BER performance. Subsection \ref{sec:Simulation results} describes the simulation setup for a real-world deployment in New York City \cite{akdeniz2014millimeter}. After that, we discuss BER performance regarding different array structures in subsection \ref{sec:BER performance analysis}. 

\vspace{-2ex}

\subsection{Simulation setup}\label{sec:Simulation results}
\vspace{-1ex}

We consider a dense urban area of New York City in which the distance between the Tx and Rx is large enough that LoS link is blocked, and the channel model is given as (\ref{channel model}) \cite{akdeniz2014millimeter}.
Parameters of considered scenario in this section are given in Table \ref{tab:System parameters}. It is considered that both the Tx and the Rx are equipped with the same antenna geometry/size ($N = N_t = N_r$)\footnote{In this paper, we investigate half-duplex (HD) link with the same array geometry in both terminals. In an HD scenario, a terminal is a transmitter at a specific time slot, while in the next time slot, it can receive data and serve as a receiver; hence, this assumption is reasonable. In a full-duplex (FD) scenario, we can adopt different array structure combinations at Tx/Rx.}. Regarding ULA, URA, and UCA, the separation between antenna elements is $\frac{\lambda}{2}$; consequently, in the case of UCA, the radius of the antenna array is $R = \frac{N\lambda}{4 \pi}$. For the CCA, the structure is slightly different because the antenna elements are not uniformly distributed. We assume a CCA structure with optimized ring spacing and number of elements in each ring as introduced in \cite{haupt2008optimized}. The radius and number of elements of the $n_{c'}$th ring is assumed to be $R_{n_{c'}} \in \{ 0.76 \lambda, 1.36 \lambda, 2.09 \lambda, 2.99 \lambda \}$ and $N_{n_{c'}} \in \{ 9, 17, 25, 31 \}$, respectively \cite{haupt2008optimized}. With this type of structure, $82$ antenna elements can be arranged in $4$ rings. 
We define path loss in dB as follows \cite{akdeniz2014millimeter}: \vspace{-1.5ex}
\begin{equation}
    PL(d) = a + 10 b \log_{10}(d) + \xi \ \mathrm{[dB]},
    \vspace{-1.5ex}
\end{equation}
where $\xi$ is the effect of large-scale shadow fading and it has a normal distribution with zero mean and standard deviation $\sigma_{\xi}$, i.e., $\xi \sim \mathcal{N}(0,\sigma_{\xi}^2)$.

\begin{figure}
    \centering
    \includegraphics[scale = 0.3]{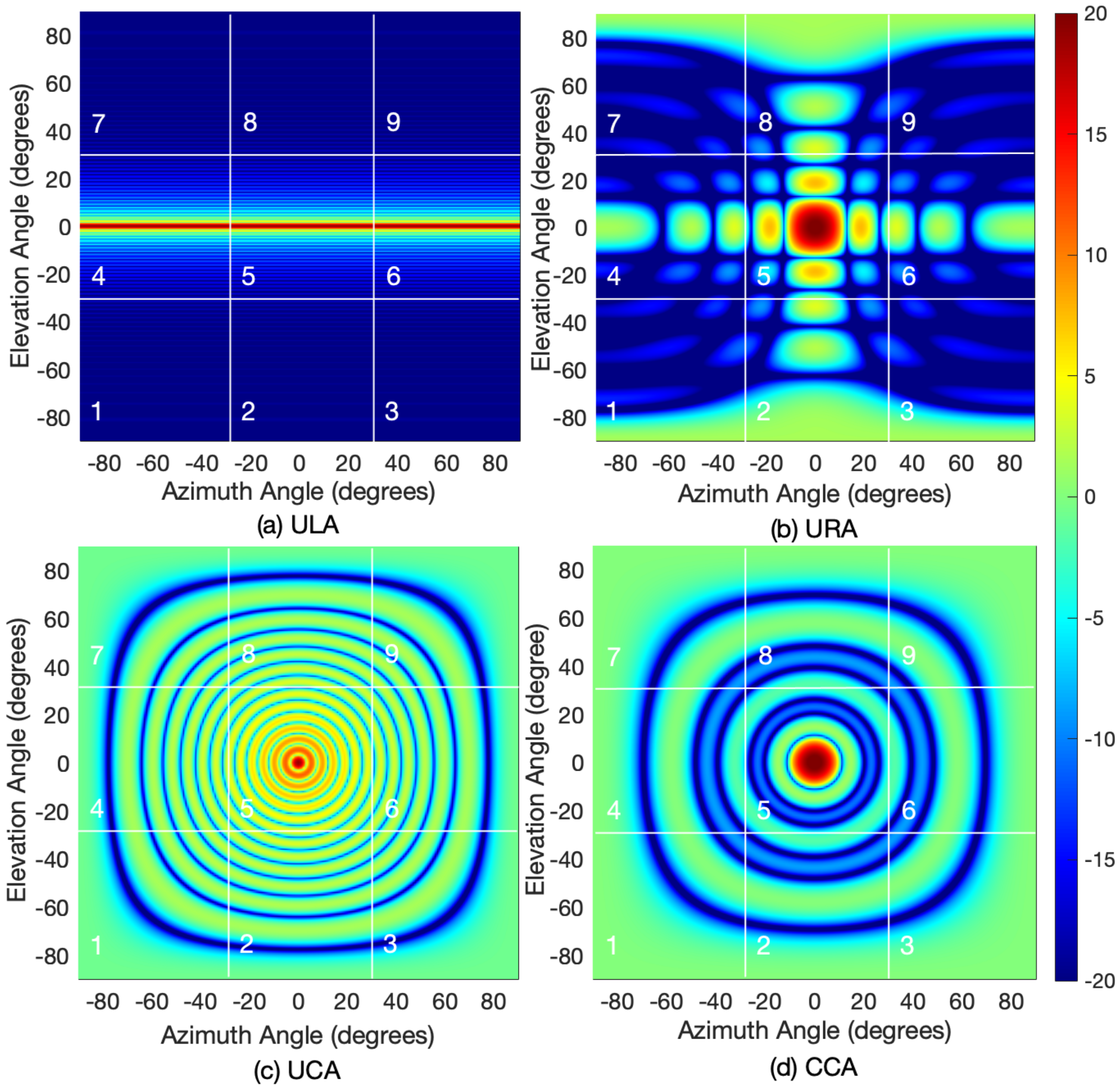}
    \caption{Antenna directivity patterns for different array architectures.}
    \label{3D_DirectivityPattern_30_60} 
\end{figure}

\vspace{-1.7ex}
\subsection{BER performance}\label{sec:BER performance analysis}
\vspace{-1ex}
This subsection demonstrates the BER comparison for different antenna array structures by presenting the Monte Carlo simulation results of BER performance. 
For the sake of analysis, we define a new radiation characteristic named side lobe directivity (SLD) which states the directivity of a side lobe emitted by an antenna array. Afterward, taking the geometrical mean of SLD gives us the average SLD (ASLD), which helps us to investigate the behavior of different array structures on a CIM-based system. Hence, we can define ASLD as $\mathrm{ASLD} = \sqrt[m]{\prod_{i = 1}^{m} \mathrm{SLD}}$, where $m$ is the number of selected SLDs\footnote{$m$ should be large enough to give us an acceptable measure of ASLD.}. 
The following explains BER performance using array characteristics, i.e., directivity, HPBW, and ASLD. In order to have a detailed look at the aforementioned array characteristics, we provide numerical values in Table \ref{tab:Array charactristics} and the corresponding antenna array directivity patterns in Fig. \ref{3D_DirectivityPattern_30_60}.

\begin{figure}
    \centering
    \includegraphics[scale = 0.32]{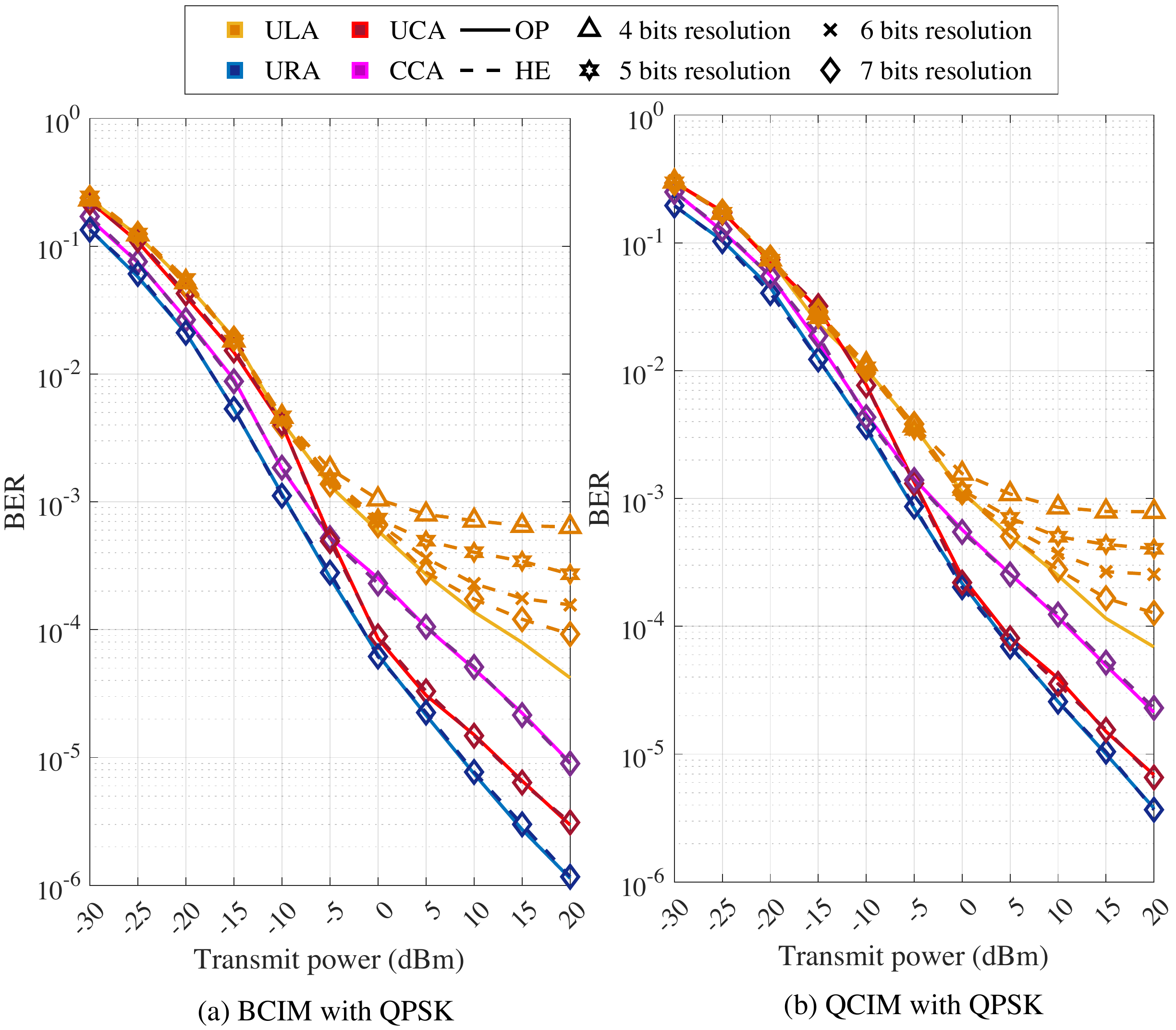}
    \caption{BER comparison of different array structures.}
    \label{fig:BER_Analysis}
\end{figure}

Fig. \ref{fig:BER_Analysis} illustrates BER performance for two different signalings, i.e., (a) binary CIM (BCIM) with QPSK and (b) quadrature CIM (QCIM) with QPSK. As expected, BER improves from linear arrays to 2D arrays due to the ability of 2D arrays to illuminate azimuth and elevation dimensions simultaneously.
Regarding 2D arrays, URA has the best BER performance. 
To explain this superiority, we notice the ASLD mentioned in Table \ref{tab:Array charactristics}. Among the 2D arrays, URA has provided significantly lower ASLD. At the Tx, lower side lobes result in less power leakage into the non-intended directions, which in turn reduces inter-cluster interference at the receiver. At the Rx, lower side lobes reduce collected interference from the non-intended indexed clusters. Fig. \ref{3D_DirectivityPattern_30_60} illustrates the side lobes effect on the non-intended angles for different array structures.

To explain the effect of side lobes, let us give an example of the studied CIM system. First, we divide each pattern in Fig. \ref{3D_DirectivityPattern_30_60} into 9 areas. Without loss of generality, let us assume the Tx is going to transmit a signal toward a cluster located in origin; therefore, the main lobe emitted by the antenna array is steered toward the origin, as depicted in Fig. \ref{3D_DirectivityPattern_30_60}. Hereupon, we call area 5 as the intended area and all the other areas as interference areas. Under such circumstances, any power leakage (signal collection) to (from) the interference areas at the Tx (Rx) contributes to performance degradation by leveraging more inter-cluster interference. As is depicted in Fig. \ref{3D_DirectivityPattern_30_60}, UCA and CCA have higher side lobes directivity compared with URA in the interference areas. More specifically, URA has extremely low directivity side lobes in areas $1$, $3$, $7$, and $9$, i.e., lower than $-15$ dB, while this value for UCA and CCA is around $0$ dB. Assume that the CIM algorithm indexes some clusters in the mentioned areas; hence, UCA and CCA entail more interference in the system compared with URA. This example explains the superiority of URA over UCA and CCA well. 

In comparison between UCA and CCA, CCA performs better than UCA in low transmit powers. This behavior can be explained by comparing of directivity and ASLD between two arrays. As expressed in Table \ref{tab:Array charactristics}, CCA has higher directivity and less ASLD than the UCA, which has the dominant effect on the BER performance and results in better performance of CCA in low transmit powers. However, with increasing transmit power, UCA performs better, as illustrated in Fig. \ref{fig:BER_Analysis}. Here, the effect of narrower beam width of UCA becomes dominant with increasing transmit power and causes better BER performance. As is provided in Table \ref{tab:Array charactristics}, the HPBW of UCA is approximately three times less than CCA.

HE analog network with implementing FPSs is adopted to reduce cost and energy consumption. As it is shown in Fig. \ref{fig:BER_Analysis}, by adopting the HE structure with only $7$ bits resolution, i.e., $8$ quantized fixed phase shifters, we can achieve the same BER performance with OP structure for 2D antenna arrays. This is while, to implement OP analog network, we need to adopt $N_t$ SPSs at the Tx and $N_{RF} \times N_r$ SPSs at the Rx. Knowing that each SPS costs around a hundred US dollars and consumes around $50$ milliwatts of power shows the superiority of implementing the HE structure well \cite{yu2018hardware}. When we increase the resolution bits, the HE BER performance can approach the OP structure more. We also depict the BER performance for different resolution bits, i.e., $4$, $5$, $6$, and $7$, for the ULA structure in Fig. \ref{fig:BER_Analysis}. As is expected, the BER performance of the HE structure approaches the OP structure with increasing resolution bits (number of adopted FPSs). It is worth mentioning that BER performance has an error floor when the HE structure is adopted. However, increasing the resolution of the HE structure reduces the error floor because the detector can resolve the symbols more precisely.  

\begin{table}
\scriptsize
    \centering
    \caption{Array characteristics for different array structures.}
    \vspace{-2ex}
    \label{tab:Array charactristics}
    \begin{tabular}{c c c c}
        \toprule
        \multirow{3}{*}{} & \multirow{3}{*}{\textbf{Parameter}} & \multicolumn{2}{c}{\textbf{Value}} \\ \cmidrule{3-4} 
        & & \textit{Steered at} & \textit{Steered at} \\
        & & \textit{$0^\circ$ Az; $0^\circ$ El} & \textit{$15^\circ$ Az; $30^\circ$ El}\\
        \toprule
        \multirow{3}{*}{\rotatebox[origin=c]{90}{\textbf{ULA}}}& \textit{Directivity (dBi)} & $19.14$ & $19.14$ \\
        & \textit{HPBW} & $360^\circ$ Az; $1.24^\circ$ El & $360^\circ$ Az; $1.28^\circ$ El\\
        & \textit{ASLD (dB)} & $-23.52$ & $-18.94$ \\
        \midrule
        \multirow{3}{*}{\rotatebox[origin=c]{90}{\textbf{URA}}}& \textit{Directivity (dBi)} & $20.67$ & $19.97$ \\
        & \textit{HPBW} & $11.34^\circ$ Az; $11.34^\circ$ El & $13.56^\circ$ Az; $13.00^\circ$ El\\
        & \textit{ASLD (dB)} & $-16.52$ & $-18.36$\\ 
        \midrule
        \multirow{3}{*}{\rotatebox[origin=c]{90}{\textbf{UCA}}}& \textit{Directivity (dBi)} & $19.32$ & $19.16$  \\
        & \textit{HPBW} & $3.16^\circ$ Az; $3.16^\circ$ El & $3.76^\circ$ Az; $3.59^\circ$ El \\
        & \textit{ASLD (dB)} & $-3.90$ & $-4.42$ \\
        \midrule
        \multirow{3}{*}{\rotatebox[origin=c]{90}{\textbf{CCA}}}& \textit{Directivity (dBi)} & $21.13$ & $19.72$ \\
        & \textit{HPBW} & $9.20^\circ$ Az; $9.20^\circ$ El & $11.00^\circ$ Az; $10.51^\circ$ El \\
        & \textit{ASLD (dB)} & $-5.91$ & $-6.63$\\ 
        \bottomrule
    \end{tabular}
\end{table}


\vspace{-2ex}

\section{Conclusion}\label{Sec: Conclusion}
\vspace{-1ex}
This paper has investigated the effect of different array structures on CIM-enabled mMIMO mmWave communications systems. We have studied the different array characteristics, i.e., directivity, HPBW, and ASLD, to show their effect on the BER performance of the CIM-enabled system. Our illustrative results revealed that URA has the best BER performance among its counterparts, thanks to the lower side lobes. It has been shown that the higher side lobes directivity results in stronger inter-cluster interference, which is detrimental to the CIM-based systems. We have also proposed an algorithm for the HE analog network to decrease cost and energy consumption significantly. It is shown that the HE structure can perform similarly to the OP structure by adopting a few numbers of FPS. As a potential direction for future work, an intelligent CIM algorithm can be designed to intelligently index clusters and limit the effect of inter-cluster interference by considering the antenna array directivity pattern.

\vspace{-1ex}

\bibliographystyle{ieeetr}
\bibliography{reference.bib}

\end{document}